\documentclass[aps,prb,twocolumn,superscriptaddress,showpacs,amsmath,amssymb,footinbib,longbibliography]{revtex4-2}
\usepackage[english]{babel}
\usepackage{amssymb,amsbsy,amsmath}
\usepackage{graphicx}% Include figure files
\usepackage{dcolumn}% Align table columns on decimal point
\usepackage{bm}% bold math
\usepackage{subfigure}
\usepackage{color}
\usepackage{textcomp}
\usepackage[colorlinks,bookmarks=false,citecolor=blue,linkcolor=red,urlcolor=blue]{hyperref}
\usepackage[english]{babel}
\newcommand{\op}[1]{%
    \fontdimen12\textfont3=2pt\fontdimen12\scriptfont3=1.4pt%
    \!\null\mathop{\vphantom{#1}\smash{#1}}\limits_{\sim}\null\!}

\newcommand{\figref}[1]{Fig.~\protect\ref{#1}}

\def\ket#1{\, | \, {#1} \, \rangle}
\newcommand{\braket}[2]{\langle \, {#1} \, | \, {#2} \, \rangle}

\begin{document}
\title{Thermodynamics of the spin-half square-kagome lattice antiferromagnet}

\author{Johannes Richter}
\email{Johannes.Richter@physik.uni-magdeburg.de}
\affiliation{Institut f\"ur Physik, Universit\"at Magdeburg, P.O. Box 4120, D-39016 Magdeburg, Germany}
\affiliation{Max-Planck-Institut f\"{u}r Physik Komplexer Systeme,
        N\"{o}thnitzer Stra{\ss}e 38, D-01187 Dresden, Germany}
\author{Oleg Derzhko}
\email{derzhko@icmp.lviv.ua}
\affiliation{Institute for Condensed Matter Physics,
          National Academy of Sciences of Ukraine,
          Svientsitskii Street 1, 79011 L'viv, Ukraine}
\author{J\"urgen Schnack}
\email{jschnack@uni-bielefeld.de}
\affiliation{Fakult\"at f\"ur Physik, Universit\"at Bielefeld, Postfach 100131, D-33501 Bielefeld, Germany}

\date{\today}

\begin{abstract}
Over the last decade, the interest in the spin-$1/2$  Heisenberg antiferromagnet
(HAF) on the square-kagome (also called shuriken) lattice has been growing
as a model system of quantum magnetism with a quantum paramagnetic ground
state, flat-band physics near the saturation field, and quantum scars.
A further motivation to study this model comes from the recent discovery of
a gapless spin liquid in the square-kagome magnet
KCu$_6$AlBiO$_4$(SO$_4$)$_5$Cl
[{\it M. Fujihala et al., Nat. Commun. \textbf{11}, 3429 (2020)}].
Here, we present large-scale numerical investigations of the specific heat
$C(T)$,
the entropy $S(T)$ as well as the susceptibility $\chi(T)$ by means of the
finite-temperature Lanczos method for system sizes of $N=18,24,30,36,42,48$, and 
$N=54$.  We find that the specific heat
exhibits a low-temperature shoulder below
the major maximum which can be attributed to low-lying singlet
excitations filling the singlet-triplet gap, which is significantly
larger than the singlet-singlet gap.
This observation is further supported by the behavior of the
entropy $S(T)$, where a change in curvature is present 
just at about $T/J=0.2$, the same temperature where the  shoulder in $C$ sets in.
For the susceptibility the low-lying singlet excitations are irrelevant, and 
the singlet-triplet gap 
leads to an exponentially activated low-temperature behavior.
The maximum in $\chi(T)$ is found at a pretty low temperature  $T_{\rm max}/J=0.146$
(for $N=42$) compared to $T_{\rm max}/J=0.935$  for the
unfrustrated square-lattice HAF signaling the crucial role of
frustration also for the susceptibility. We find a striking
similarity of our square-kagome data with the corresponding ones for
the kagome HAF down to very low $T$. 
The magnetization process featuring plateaus and jumps and
the field dependence of the specific heat that exhibits
characteristic peculiarities attributed to the existence of a
flat one-magnon band are as well discussed.
\end{abstract}


\maketitle

%%%%%%%%%%%%%%%%%%%%%%%%%%%%%%%%%%%%%%%%%%%%%%%%%%%%%%%%%%%%%%%%%%%%%%%%
\section{Introduction}

The spin-$1/2$ Heisenberg antiferromagnet (HAF) 
exhibits N\'{e}el semi-classical magnetic long-range order on
most of the two-dimensional (2D) lattices. Among the eleven 2D
Archimedean tilings  \cite{def_Archi,gruenbaum},
only the celebrated kagome lattice and the so-called star lattice
have non-magnetic spin-liquid ground states
\cite{LNP645,Bal:N10,farnell2014quantum,IBS:PRB13,SaB:RPP17,Orus_star_2018,PhysRevB.98.224402}.
Both lattices have low coordination number $z$, and they are highly
frustrated.
A non-Archimedean lattice without magnetic long-range order is the square-kagome
(sometimes also called shuriken) lattice
\cite{Sidd2001,richter2004squago,richter2009squago,Sakai2013}. Similar
to the kagome lattice 
it is a 2D tiling of corner-sharing triangles, i.e. the classical ground state
of the square-kagome  Heisenberg antiferromagnet (SKHAF) 
is highly degenerated. 
There are two non-equivalent sites A and B and nearest-neighbor bonds $J$ and $J'$, see the inset in Fig.~\ref{fig_W}.
In our paper we will focus on the balanced situation $J=J'=1$ 
which is most
similar to the kagome HAF.  
The corresponding Hamiltonian augmented with a Zeeman term is given by 
\begin{eqnarray}
\label{Ham}
\op{H}
&=&
J \sum_{<i,j>}\;
\op{\vec{s}}_i \cdot \op{\vec{s}}_j
+
g \mu_B\, B\,
\sum_{i}\;
\op{s}^z_i
\ ,
\end{eqnarray}
where
quantum mechanical operators are marked by a tilde.

Over the last decade, the interest in the spin-$1/2$ SKHAF has been growing
as a model system of quantum magnetism with a quantum paramagnetic ground
state 
as well as flat-band physics near the saturation field and quantum scars 
\cite{Sakai2013,Rousochatzakis2013,derzhko2014square,Rousochatzakis2015,richter2004-spin-peierls,Sakai2015,DRM:IJMP15,Hasegawa2018,Morita2018,Lugan2019,McClarty2020,PhysRevB.102.241115,Iqbal2021}.
It is expected that the low-energy physics of the 
SKHAF is just as fascinating as for the related kagome HAF.
While it is not questioned that the strong frustration prevents ground-state
magnetic ordering, the specific nature of the quantum ground state is under debate.
Thus, a recent Schwinger-boson approach leads to a topological spin-liquid
ground state  
with weak nematicity and a small gap \cite{Lugan2019}, whereas  
a variational Monte-Carlo study is in favor of a gapped pinwheel valence-bond
crystal state \cite{Iqbal2021}. 
Previous exact-diagonalization studies seem to also suggest a finite
singlet-triplet gap (spin gap)
\cite{richter2009squago,Sakai2013}, though this issue seems as yet
unresolved. However and again similar to the kagome HAF
\cite{Lech:1997,Waldtmann1998,LSM:PRB19}, there are many low-lying singlet excitations
filling the singlet-triplet spin gap \cite{richter2009squago}.
In addition to the zero-field properties, the magnetization process is of
particular interest. Prominent features of the magnetization curve $M(B)$ are a
wide plateau at $1/3$ of the saturation magnetization $M_{\rm sat}$ and a jump to saturation
with a preceding plateau at $(2/3)M_{\rm sat}$ \cite{richter2009squago,Sakai2013}.
A further motivation to study this model comes from recent
experimental investigations. A gapless spin liquid was dicovered
in the square-kagome magnet
KCu$_6$AlBiO$_4$(SO$_4$)$_5$Cl \cite{FMM:NC20,Vasiliev2022} which is, however, not
a well-balanced (i.e.
$J \ne J'$) SKHAF; rather in this compound three different exchange couplings
are relevant.
For {Na$_6$Cu$_7$BiO$_4$(PO$_4$)$_4$[Cl,(OH)]$_3$} measurements
of the magnetization as well as of the heat capacity indicate
the absence of long-range order which might be a signal of spin
liquid behavior \cite{YSK:IC21,Vasiliev2022}.

While almost all previous studies of the SKHAF are focused on ground-state
properties,
the  thermodynamics of the model is much less investigated.
Only, in the early paper Ref.~\cite{tomczak2003specific} the specific heat
was calculated by a simple renormalization group approach.
In our paper we want to fill this gap of missing finite-temperature studies
of the spin-$1/2$ SKHAF 
by  large-scale numerical simulations of finite lattices of up to $N=54$
sites (see Fig.~\ref{fig_lat} in the
Appendix) by means of full exact diagonalization (ED) \cite{Lauchli_ED_2011}
and of the finite-temperature Lanczos method
(FTLM)
\cite{JaP:PRB94,JaP:AP00,ScW:EPJB10,ScT:PR17,PrK:PRB18,kago42,PRL_mag_cryst,MoT:A20,Accuracy2020}.
We will use the well investigated kagome HAF 
as reference system when discussing the data of the  SKHAF.

The very existence of an excitation gap is crucial for low-temperature thermodynamics
at low temperatures $T$. Thus, a singlet-triplet gap leads to an
exponentially activated low-temperature behavior of the  
susceptibility, whereas low-lying non-magnetic singlet excitations are relevant for
the specific heat.  
As indicated by previous Lanczos data for finite lattices up to $N=36$ sites
\cite{richter2009squago} 
similarly to the kagome HAF we may expect a small singlet-triplet gap
filled with a noticeable  number of singlet states.

\textcolor{black}{
It is appropriate to mention here, that for smaller cluster sizes some related data have 
already been reported in previous works. 
That concerns (i) 
the zero-temperature magnetzation curves shown in Sec.~\ref{field}, cf.
Refs.~\cite{richter2009squago,Sakai2013}.
However, here we add some new data for larger lattice sizes of $N=48$ and  
$N=54$ sites and we present the finite-size dependence of the widths of the 
magnetization plateaus.   
Furthermore, (ii) it concerns
the specific heat in strong magnetic fields near saturation.
In Ref.~\cite{richter2009squago} one can find related specific-heat data for much smaller sizes of
$N=18$ and $N=24$.
In our paper we show specific-heat data up to $N=54$. }

The paper is organized as follow. In Section \ref{sec-2} we
introduce our numerical scheme, in
Section~\ref{sec-3} we present our results for the SKHAF and compare them
with corresponding data for 
kagome HAF. 
In the last Section~\ref{sec-4}
we discuss and summarize our findings.
For convenience we show the  
finite lattices considered here
in an  appendix.

%%%%%%%%%%%%%%%%%%%%%%%%%%%%%%%%%%%%%%%%%%%%%%%%%%%%%%%%%%%%%%%%%%%%%%%%
\section{Calculational scheme}
\label{sec-2}

The investigated spin system is modeled by the spin-$1/2$
Heisenberg Hamiltonian given in Eq.~(\ref{Ham}).
The complete spectrum for the spin-half system can be calculated only for
the smallest finite lattice of $N=18$ spins.
For larger systems we perform full diagonalization in high sectors of
magnetization $M$ (e.g., for $N=42$ in subspaces with $M=16,\ldots,21$). 
The obtained exact energy spectrum yields the contribution of these $M$
sectors to the partition function $Z(T,B)$.   
  
For subspaces that are not accessible by full exact
diagonalization we use the finite-temperature Lanczos method
(FTLM) 
\cite{JaP:PRB94,JaP:AP00,PrB:SSSSS13,PRE:COR17}  
which provides approximations
of thermodynamic quantities with remarkable
accuracy \cite{ScW:EPJB10,ScT:PR17,Accuracy2020}. Within the
FTLM scheme
the sum over an orthonormal basis in the  partition function is replaced by a much smaller sum over
$R$ random vectors: 
%--------------------------------------------------------
\begin{eqnarray}
\label{Z}
Z(T,B)
&\approx&
\sum_{\gamma=1}^\Gamma\;
\frac{\text{dim}({\mathcal H}(\gamma))}{R}
\sum_{\nu=1}^R\;
\sum_{n=1}^{N_L}\;
e^{-\beta \epsilon_n^{(\nu)}} |\braket{n(\nu)}{\nu}|^2
\ .
\nonumber \\[-3mm]
\end{eqnarray}
%--------------------------------------------------------
Here the $\ket{\nu}$ label  random vectors 
for each symmetry-related orthogonal subspace
${\mathcal H}(\gamma)$ of the Hilbert space, where $\gamma$
denotes  the respective 
symmetry. 
In Eq.~(\ref{Z}) the exponential of the Hamiltonian is approximated by
its spectral representation in a Krylov space spanned by the
$N_L$ Lanczos vectors starting from the respective random vector
$\ket{\nu}$, where
$\ket{n(\nu)}$ is the $n$-th eigenvector of $\op{H}$ in
this Krylov space. 
This method is known to provide  
accurate data for
typical observables
such as magnetization, uniform magnetic susceptibility  and
specific heat, see, e.g. \cite{Accuracy2020}.

Naturally, we take into account the commutation of the Hamiltonian $\op{H}$ with the
$z$-component of the total spin  $\op{S}_z=\sum_{i}\;
\op{s}^z_i$ to decompose the full Hilbert space 
into much smaller
orthogonal subspaces which can be labeled by the magnetization
 $M=\langle \op{S}_z \rangle$.
For a further 
decomposition of the Hilbert space by employing lattice symmetries 
we use J\"org Schulenburg's publicly available {\it spinpack} package
\cite{spin:256,richter2010spin}. 

For the number of random vectors $R$ used to approximate the partition function 
we chose at least $R=20$ which is sufficiently large to ensure very accurate FTLM data,
cf. Refs.~\cite{kago42,PRL_mag_cryst}. However, for $N=42$, where the largest
Hilbert-space dimension is $N_H=7.34 \times 10^{10}$, 
we used $R=5$ in the subspaces of $M=0$ (that contains the ground state and the
lowest energy levels), $R=2$ for $M=1$ and $M=2$,
and then $R=10$ for $2 < M < 16$.
This strategy was used already in Ref.~\cite{kago42}, where
also the accuracy of this approach was evaluated.

%%%%%%%%%%%%%%%%%%%%%%%%%%%%%%%%%%%%%%%%%%%%%%%%%%%%%%%%%%%%%%%%%%%%%%%%
\section{The square-kagome lattice antiferromagnet}
\label{sec-3}

In what follows we discuss the Wilson ratio,  the density of
states, the specific heat, the uniform susceptibility, 
the entropy and the magnetization process.

%%%%%%%%%%%%%%%%%%%%%%%%%%%%%%%%%%%%%%%%%%%%%%%%%%%%%%%%%%%%%%%%%%%%%%%%
\subsection{Zero-field properties}
\label{zero_B}

First we study the modified Wilson ratio \cite{Wilson_PRB2020,PRR2020_Wilson} 
\begin{equation} \label{W}
P(T)=4\pi^2T \chi /(3S),
\end{equation}
where $\chi$  is the uniform magnetic susceptibility  and $S$ is the entropy.
It measures 
the ratio of the density of magnetic excitations with $M>0$ 
and the density of {\it all} excitations  including singlet  excitations with
$M=0$.
As demonstrated for the kagome HAF \cite{Wilson_PRB2020,PRR2020_Wilson},
a vanishing $P$  as temperature $T \to 0$   
is a hallmark of quantum spin-liquid ground state with dominating singlet
excitations at low $T$.
On the other hand, for quantum spin models with semi-classical magnetic ground-state
order, 
such as the square-lattice HAF the  Wilson ratio diverges as
$P(T \to 0) \propto T^\eta$, $\eta \ge 1$.

%===================    figure   =================================
\begin{figure}[ht!]
\centering
\includegraphics*[clip,width=0.95\columnwidth]{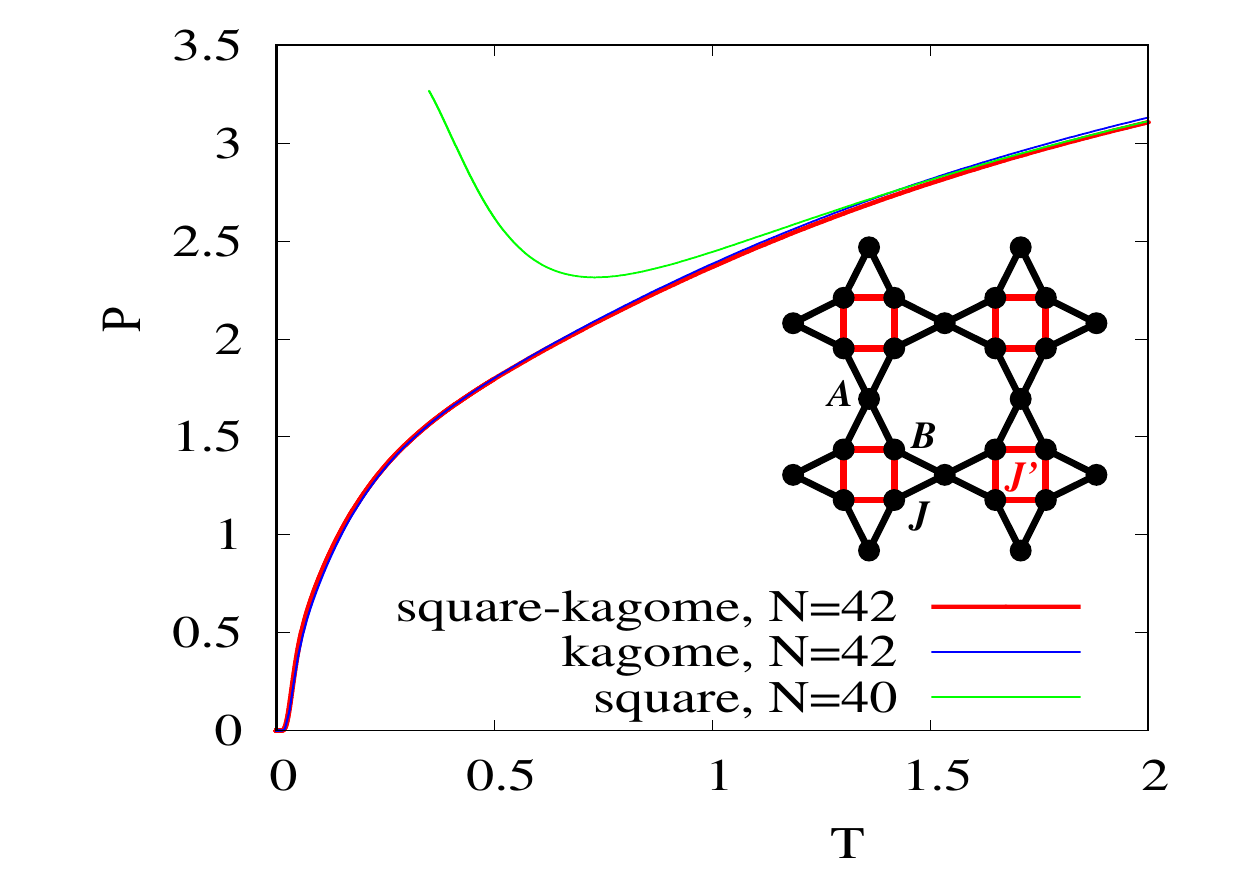}
\caption{Main panel: Modified Wilson ratio $P$, cf. Eq.~(\ref{W}),
of the spin-$1/2$ SKHAF ($N=42$) compared with respective data of the kagome
HAF
($N=42$) and square-lattice
HAF ($N=40$).
Note that the data for $N=36$ (not shown) practically  coincide with the data
for $N=42$. Note further   
that corresponding  data  for $N=36$ for the kagome and the square-lattice
HAF where previously presented in Ref.~\cite{Wilson_PRB2020}.
Inset: Sketch of the square-kagome lattice. Here A and B label the two non-equivalent sites
and $J$ and $J'$ label the two non-equivalent nearest-neighbor bonds. All
calculations are preformed for the balanced case $J'=J$.
 }
\label{fig_W}
\end{figure}
%===================    figure   =================================

In Fig.~\ref{fig_W} we show the Wilson ratio  for the $N=42$ SKHAF in comparison with
the $N=42$ kagome HAF (both with non-magnetic spin liquid ground state)  as
well as  for the $N=40$ square-lattice HAF
(with a magnetically ordered ground state).
The striking accordance of the kagome and square-kagome data is obvious
signaling the dominance of singlets as $T \to 0$. 
This behavior is in agreement with the findings reported in
Ref.~\cite{richter2009squago}  where a
noticeable number of singlets was found below the first triplet excitation.

%===================    figure   =================================
\begin{figure}[ht!]
\centering
\includegraphics*[clip,width=0.95\columnwidth]{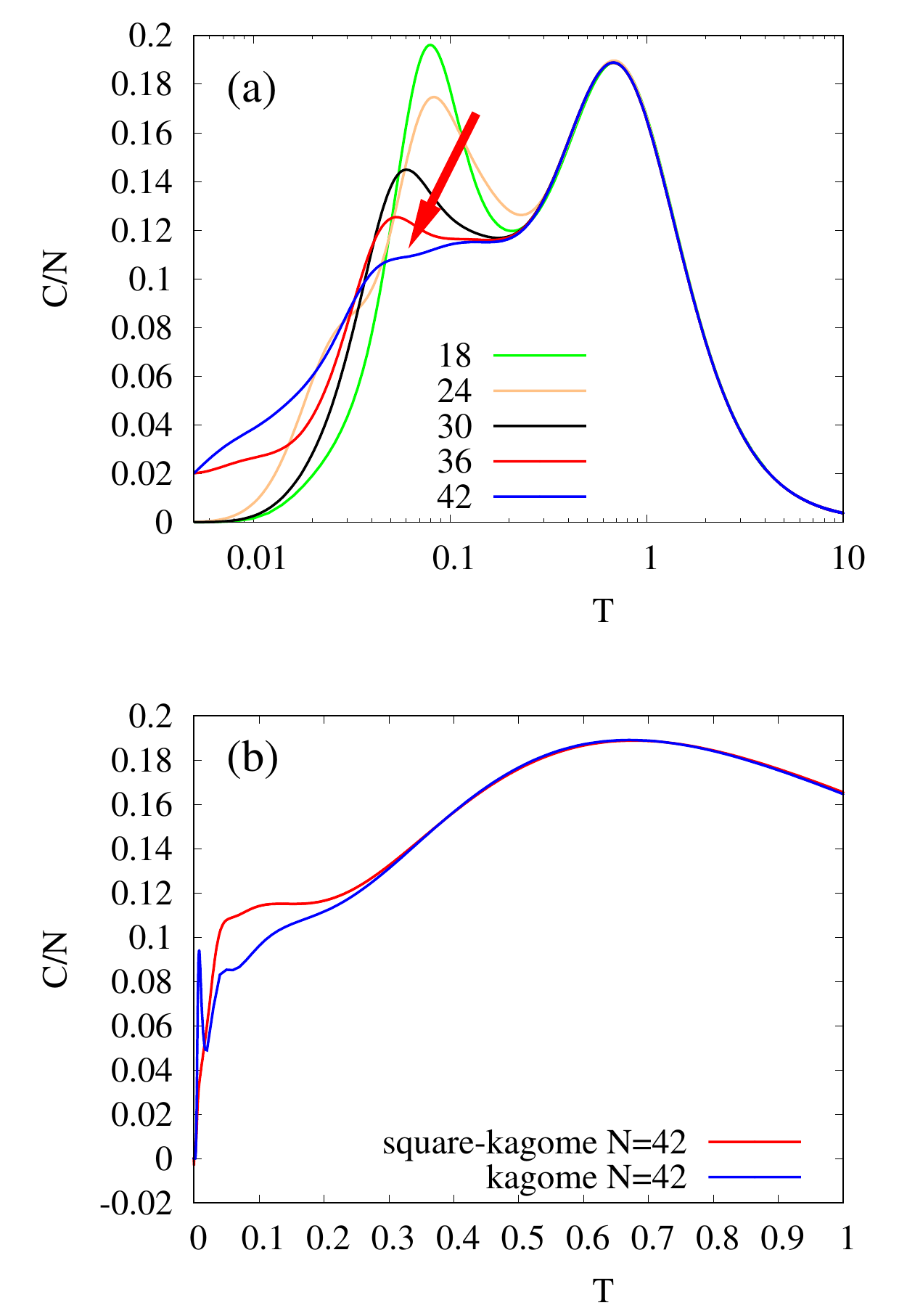}
\caption{(a) Specific heat of the SKHAF as function of temperature
 for various systems sizes (logarithmic temperature
scale). Lattice structures for $N=24,30, 36, 42$ as in
\cite{Sakai2015}, also see appendix. The red arrow shows the emergence of
the shoulder in $C(T)/N$.
(b) Specific heat of the SKHAF and the kagome HAF 
($N=42$) as function of temperature (linear
temperature scale).} 
\label{fig_C}
\end{figure}
%===================    figure  =================================

\begin{figure}[ht!]
\centering
\includegraphics*[clip,width=0.95\columnwidth]{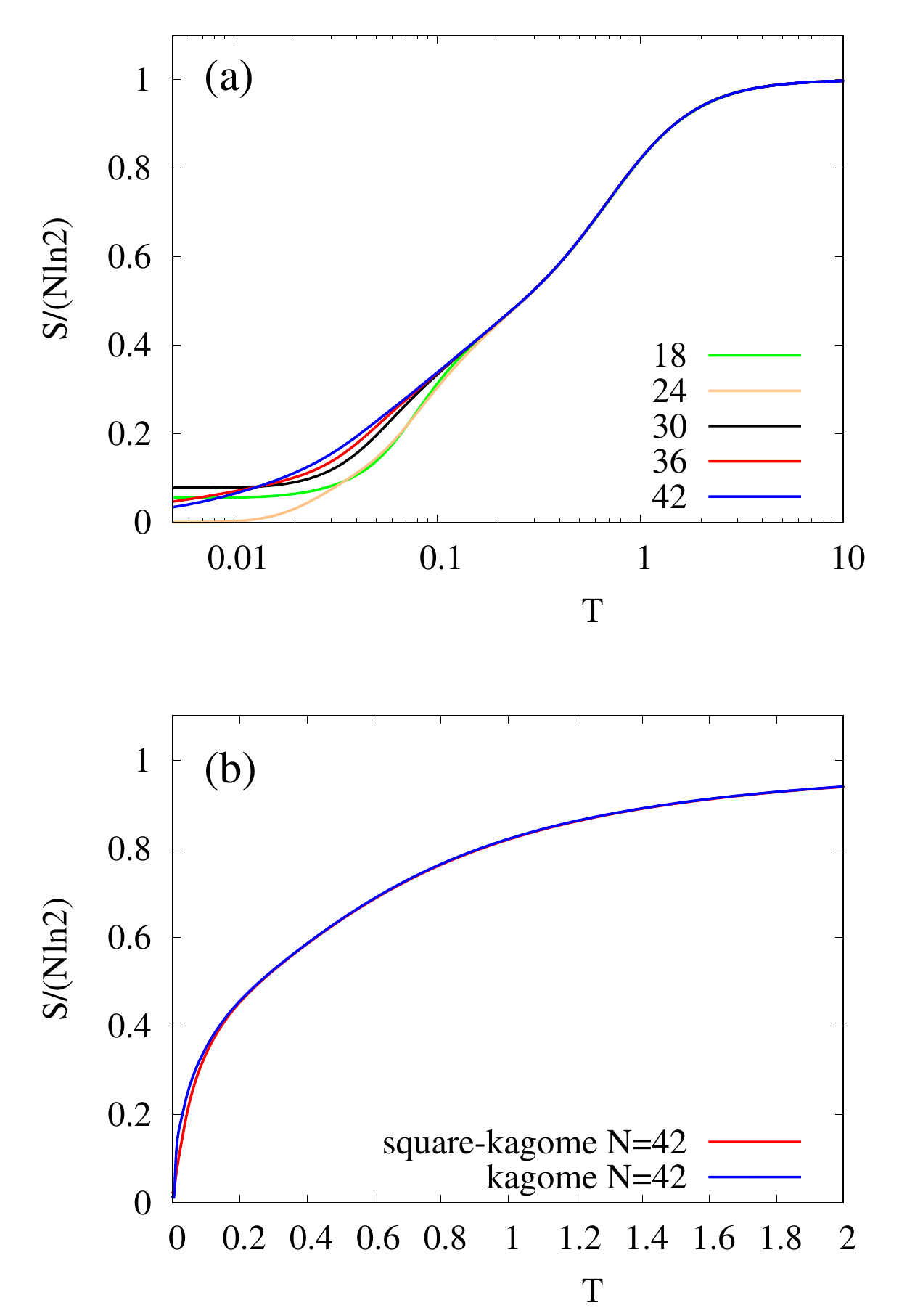}
\caption{(a) Entropy of the SKHAF as function of temperature
for various systems sizes (logarithmic temperature
scale). (b)  Entropy of the SKHAF and the kagome HAF ($N=42$) as function of temperature
(linear temperature scale).}
\label{fig_S}
\end{figure}

\begin{figure}[ht!]
\centering
\includegraphics*[clip,width=0.95\columnwidth]{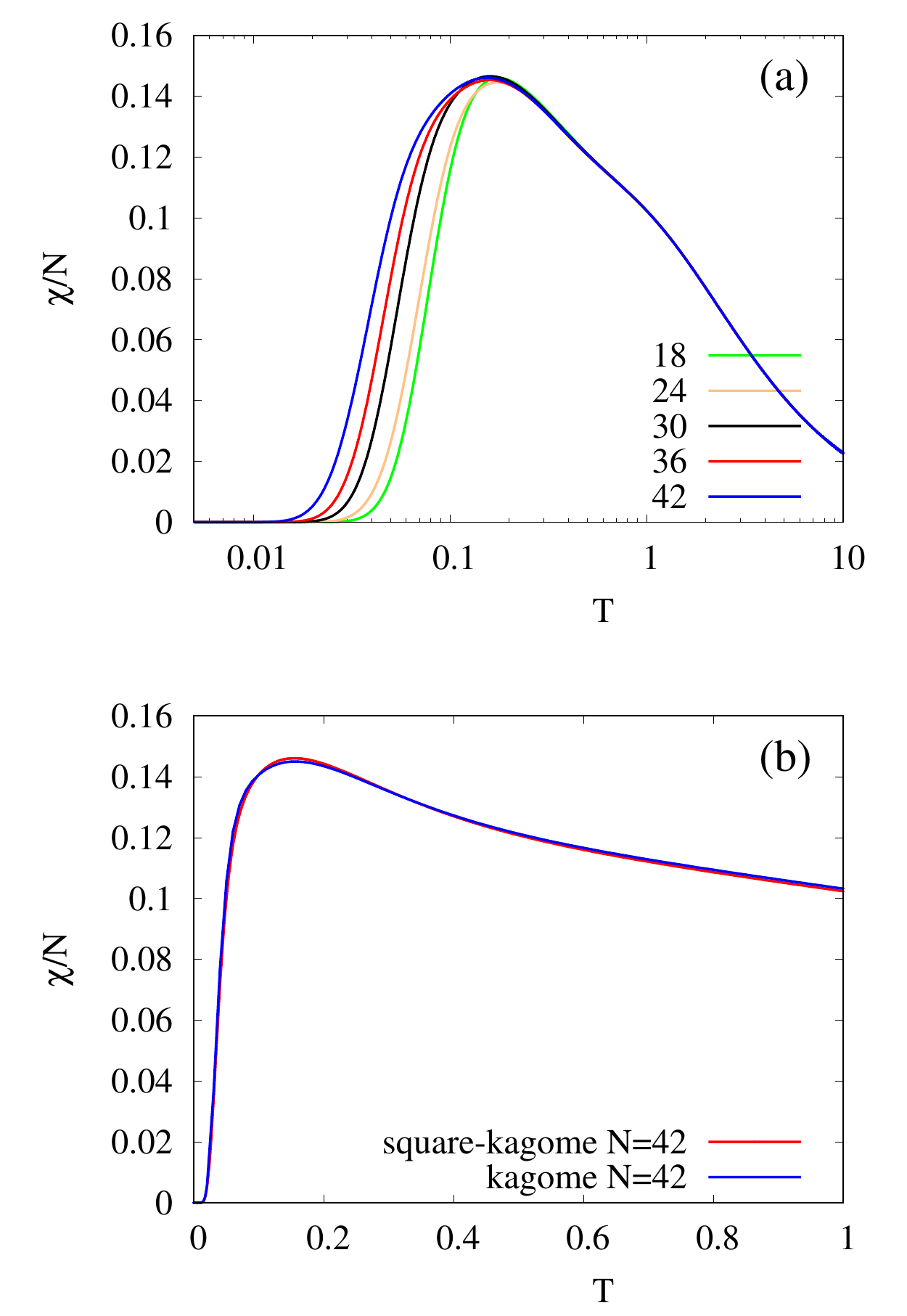}
\caption{(a) Susceptibility of the SKHAF as function of temperature
for various systems sizes (logarithmic temperature
scale). (b)  Susceptibility of the SKHAF and the kagome HAF
Heisenberg
antiferromagnet ($N=42$) as function of temperature (linear temperature scale).}
\label{fig_chi}
\end{figure}

Next we  discuss the specific heat $C(T)$, the
entropy $S(T)$ and the uniform 
susceptibility $\chi(T)$. We use a logarithmic scale for $T$ that
makes the 
low-temperature features transparent, see
Figs.~\ref{fig_C}(a), \ref{fig_S}(a), and \ref{fig_chi}(a).
In corresponding panels (b) 
we compare kagome and square-kagome results using a linear temperature scale.
The position $T_{\rm max}=0.67$ and height  $C_{\rm max}=0.189N$
of the main maximum of $C(T)$, set by the exchange 
coupling $J=1$, coincide for both models
\footnote{Temperatures
  and energies are given as multiples of the exchange coupling $J=1$, thereby omitting
  $k_B$.  $T_{\rm max}=0.67$ thus means $k_B T_{\rm max}=0.67 J$.}.

%===================    figure  =================================
\begin{figure}[ht!]
\centering
\includegraphics*[clip,width=0.95\columnwidth]{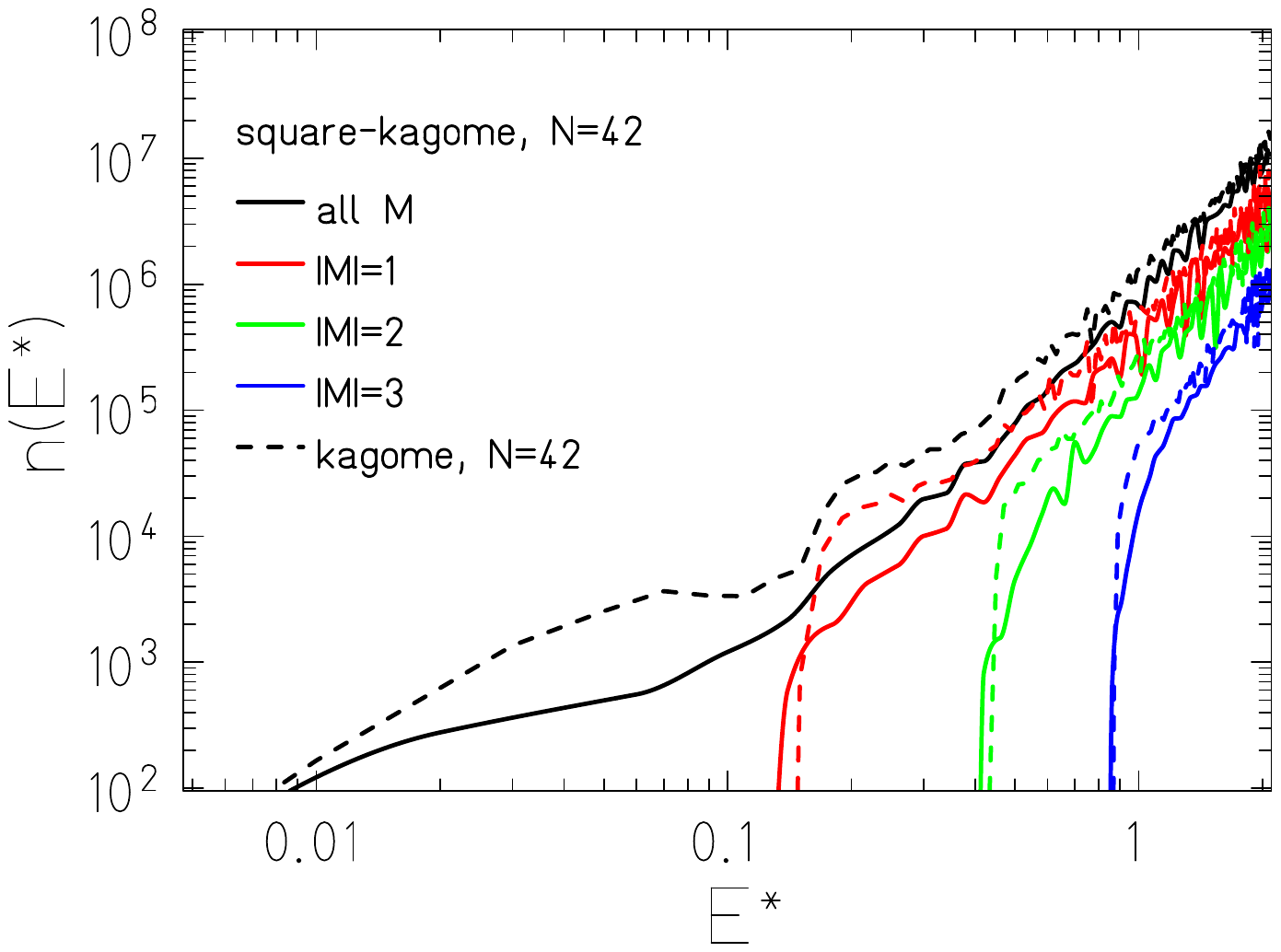}
\caption{Histogrammed density of states of the
  $N=42$ SKHAF (solid curves) and the kagome HAF (dashed curves) as a
  function of the respective excitation energy $E^*$: total
  density of states -- black, total density of states for
  $|M|=1$ -- red, for  $|M|=2$ -- green, and for $|M|=3$ -- blue
  (curves from left to right). For technical details,
  see \cite{FTLM-DOS}.
}
\label{fig_DOS}
\end{figure}
%===================    figure  =================================

%===================    figure   =================================
%===================    figure  =================================

%===================    figure   =================================
%===================    figure  =================================

Below $T \sim  0.2$ the curvature of $C(T)$ changes and a shoulder-like profile
or an additional low-temperature maximum 
appears for $0.05 \lesssim T \lesssim 0.2$. 
The very existence of such an unconventional low-temperature feature below
the main maximum   
seems to be size-independent, where  
the shrinking height of the low-$T$ maximum with growing $N$ [marked by the
red arrow in Fig.~\ref{fig_C}(a)] indicates that
there will survive 
rather a shoulder than an additional maximum in the thermodynamic limit.
At very low temperatures the singlet excitations dominate the temperature
dependence of $C$. For $N=36$ and $N=42$ the singlet-singlet gap is the smallest
and therefore $C$ remains non-zero even below $T\sim 0.01$.

In Fig.~\ref{fig_C}(b) we  compare $C(T)$ of the SKHAF and of the kagome HAF for
$N=42$. There is an almost perfect coincidence down to $T = 0.3$.
Below this temperature deviations between the curves are obvious, which
can
be attributed to subtle details in the low-energy singlet excitation
spectrum (see also the discussion of the susceptibility given below, where singlet
excitations do not play a role).
However, the shoulder-like profile is present in both systems, whereas
the sharp low-$T$ peak in $C(T)$ below the shoulder observed for the kagome HAF is absent
for the SKHAF. \textcolor{black}{Note that very recently such a shoulder of the low-temperature specific heat
has been found in the kagome quantum antiferromagnet
YCu$_3$(OH)$_6$Br$_2$[Br$_x$(OH)$_{1-x}$] \cite{PhysRevB.105.L121109}.}

To shed light on 
the relevance of low-lying  excitations of different  sectors of $M$ for the
low-temperature behavior of $C(T)$ 
we show the contributions of the different  
sectors of total magnetization $M$
to the density of states $n(E^*)$ as a function of the
respective excitation energy $E^*$ in \figref{fig_DOS}, where
$n(E^*)$ is calculated  
by histograming the Krylov space energy eigenvalues in combination
with their respective weights. 
The dominance of the singlet excitations below $E^* \sim 0.1$ is obvious.  
This observation is further
supported by the behavior of the entropy $S(T)$ 
as shown in Fig.~\ref{fig_S}(a), where a change in the curvature is present
just at about $T=0.2$. As for the specific heat finite-size effects become
visible below about  $T=0.2$. 
The comparison of the $S(T)$ profiles of the kagome HAF and SKHAF, see
Fig.~\ref{fig_S}(b), confirms again the striking similarity of both  models up
to pretty low $T$.

Next we turn to the zero-field  susceptibility $\chi$ displayed
in Fig.~\ref{fig_chi}.  
As shown in Fig.~\ref{fig_DOS} the relevant singlet-triplet gap (spin gap) is significantly larger than the singlet-singlet
gap leading to an exponentially activated low-temperature behavior.
Since the  singlet-triplet gap shrinks with growing $N$, cf.
Ref.~\cite{Sakai2013}, this feature sets in at lower $T$ as increasing $N$.
However, the question of a finite spin gap as $N \to \infty$ seems to
be not clarified so far. 
Since  the non-magnetic singlet excitations are irrelevant for the
susceptibility, the temperature profiles
of $\chi$  of the SKHAF and the kagome HAF show an excellent
agreement also below $T=0.3$, where the specific heat starts to deviate for
both models.
As already discussed in Refs.~\cite{kago42,HoA:PRB18} for the kagome HAF,
the maximum in $\chi(T)$ is at a pretty low temperature  $T_{\rm max}=0.146$
(for the $N=42$ SKHAF) compared to $T_{\rm max}=0.935$  for the
square-lattice HAF \cite{Johnston2011,SLR:PRB11}, thus 
signaling the crucial role of frustration also for the susceptibility.

%%%%%%%%%%%%%%%%%%%%%%%%%%%%%%%%%%%%%%%%%%%%%%%%%%%%%%%%%%%%%%%%%%%%%%%%
\subsection{Field-dependent properties}
\label{field}

%===================    figure   =================================
\begin{figure}[ht!]
\centering
\includegraphics*[clip,width=0.95\columnwidth]{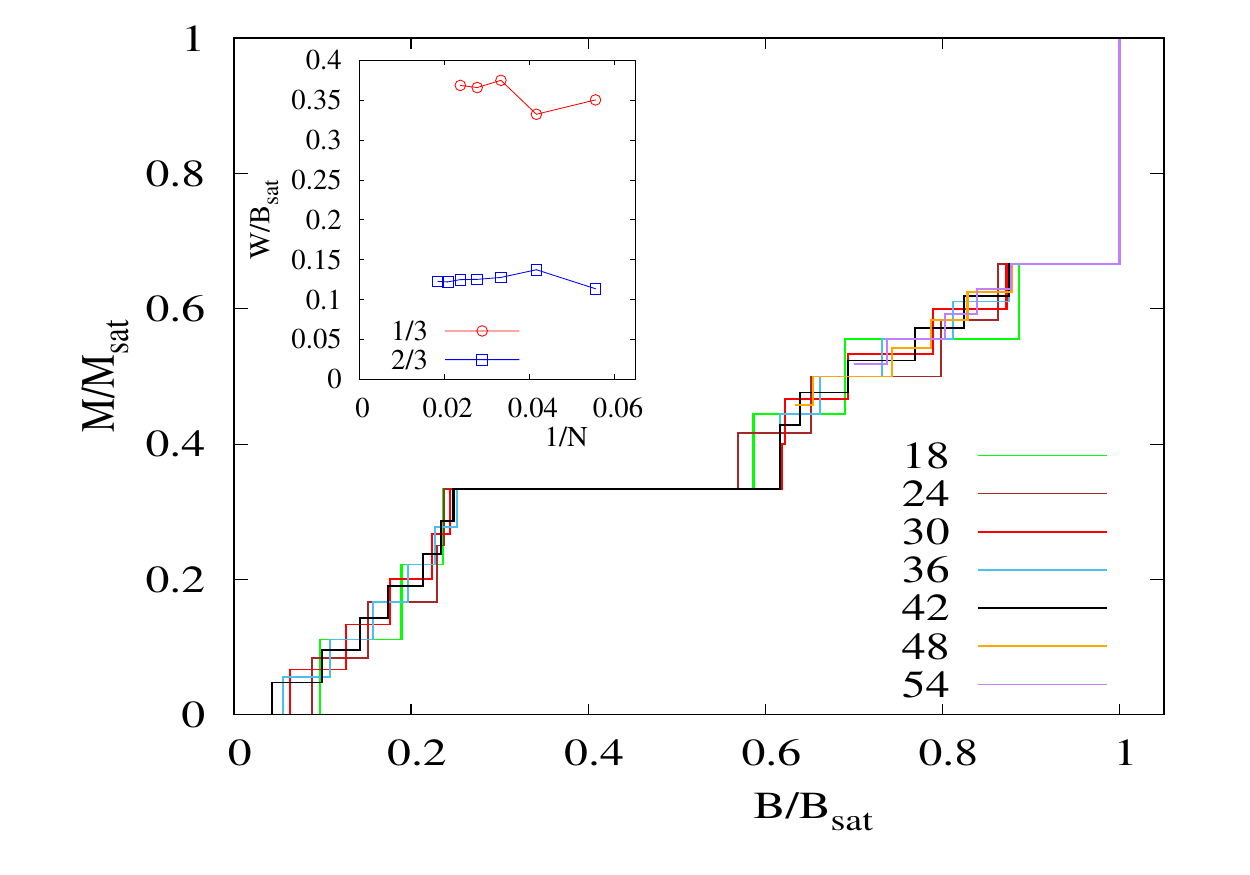}
\caption{Magnetization of the SKHAF for various lattice
  sizes and $T=0$ (main panel) and width of plateaus (inset).
 }
\label{fig_M_H}
\end{figure}
%===================    figure  =================================

%===================    figure   =================================

\begin{figure}[ht!]
\centering
\includegraphics*[clip,width=1.1\columnwidth]{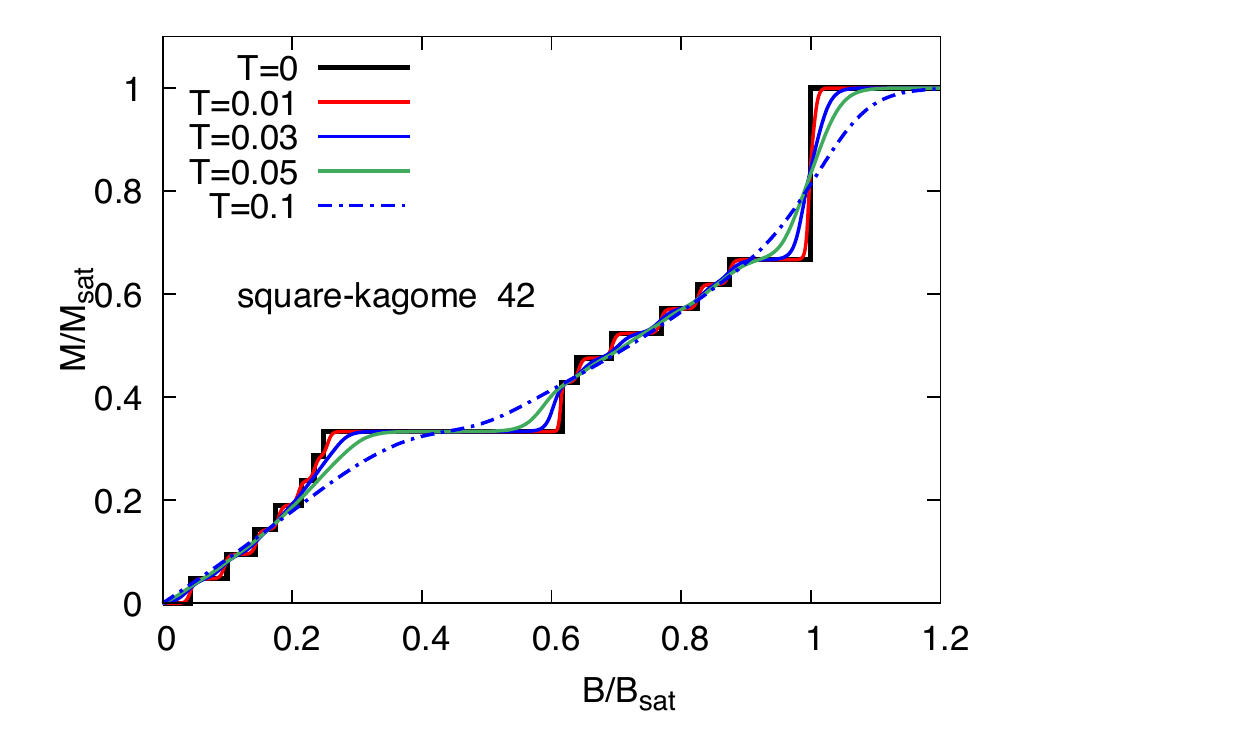}
\caption{Magnetization of the SKHAF for $N=42$ and $T\geq 0$.}
\label{fig_M_H_T}
\end{figure}
%===================    figure  =================================

%===================    figure   =================================
\begin{figure}[ht!]
\centering
\includegraphics*[clip,width=1.1\columnwidth]{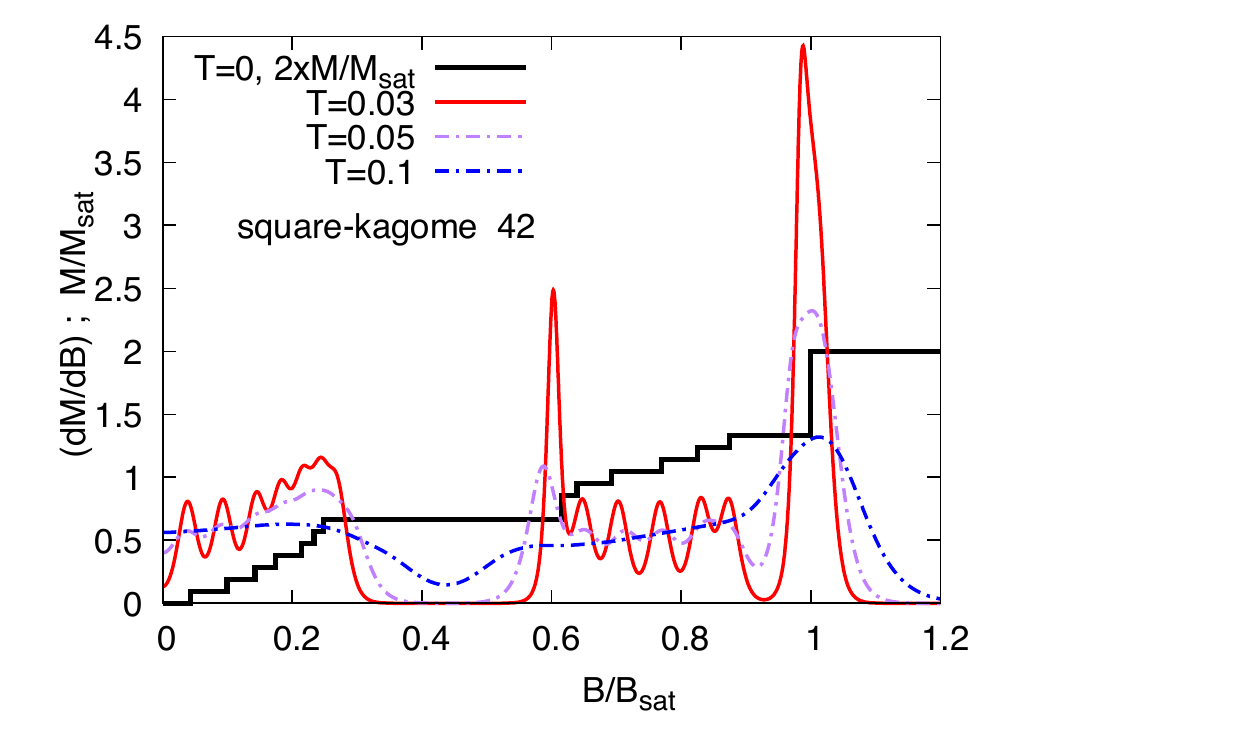}
\caption{Differential susceptibility $(dM/dB)$ of the SKHAF
  for various
  temperatures.
}
\label{fig_dM_dH_T}
\end{figure}
%===================    figure  =================================

%===================    figure   =================================
\begin{figure}[ht!]
\centering
\includegraphics*[clip,width=1.1\columnwidth]{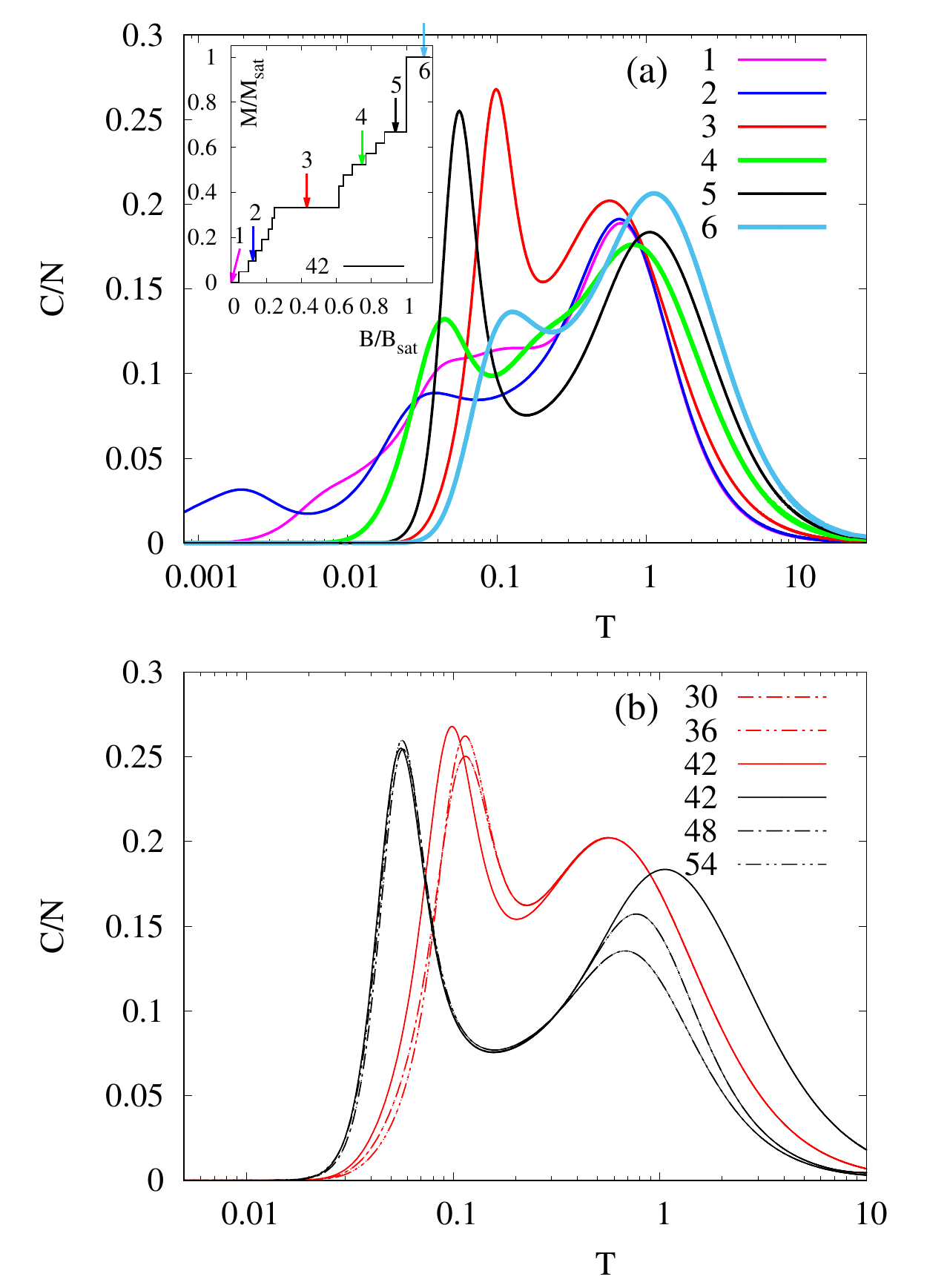}
\caption{(a) Main: Specific heat of the SKHAF for
  $N=42$ and selected values of $B$. 
Inset: $M(B)$ with arrows indicating the  values of $B$ used in the main
panel. (b) Finite-size dependence of the specific heat
for $B$ values in the middle of the $1/3$ (red lines) and $2/3$ (black
lines) plateaus. Note that for $N=48$ and $54$ lower sectors of $S_z$ are not
taken into
account in Eq.~(\ref{Z}) with the result that the upper
maximum is not correctly evaluated.  
 }
\label{fig_C_var_H}
\end{figure}
%===================    figure  =================================

%===================    figure   =================================
\begin{figure}[ht!]
\centering
\includegraphics*[clip,width=1.1\columnwidth]{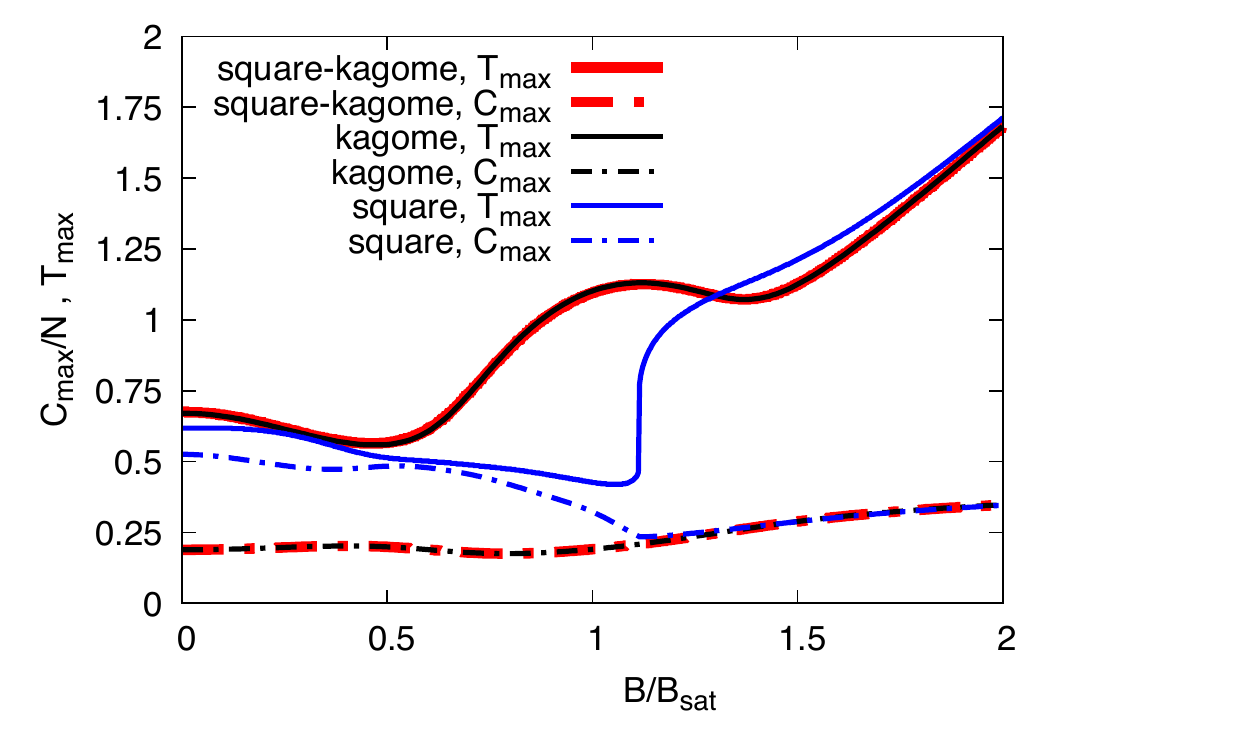}
\caption{Position $T_{\rm max}$ (solid line) and height $C_{\rm max}$
(dashed-dotted line) of the upper
main maximum in $C(T)$ in dependence on the 
magnetic field $B$ shown for the SKHAF of $N=42$ sites, the kagome HAF of $N=42$ sites
and the square-lattice  HAF of $N=40$ sites.
 }
\label{fig_Cmax_Tmax}
\end{figure}
%===================    figure  =================================

The magnetization process of strongly frustrated quantum magnets exhibits   
a number of
interesting features, such as plateaus and jumps \cite{HSR:JP04}.
We start with a brief discussion of the ground-state magnetization curve $M(B)$ of the
SKHAF, see Fig.~\ref{fig_M_H}. 
In previous studies \cite{richter2009squago,Sakai2013} $M(B)$ curves have
been reported for $N=18,24,30$ and $36$.    
Here we add
new data for $N=42$ (full $M(B)$ curve), $48$ 
($N/2 \ge M \ge 11$) and $54$ ($N/2 \ge M \ge 14$).
The saturation field $g\mu_B B_{\rm sat}=3$ is the same as for the kagome HAF.
Magnetization plateaus exist at $1/3$ and $2/3$ of the
saturation magnetization $M_{\rm sat}$ for the infinite system at $T=0$.
The widths $W$ of the plateaus are pretty large. From our data
we  estimate   $W_{1/3} \approx 0.368 B_{\rm sat}$ and
$W_{2/3} \approx 0.123 B_{\rm sat}$ (see the inset in  Fig.~\ref{fig_M_H}),
which is larger than corresponding plateau widths of the kagome HAF
\cite{NSH:NC13,CDH:PRB13}.
Moreover, there is the typical macroscopic
jump to saturation due to the presence of independent localized
multi-magnon eigenstates stemming from a flat one-magnon
band \cite{SSR:EPJB01,SHS:PRL02,derzhko2004finite,zhitomirsky2004exact,DRH:LTP07}.
Its existence is analytically proven and it does not exhibit
finite-size effects \cite{SSR:EPJB01,SHS:PRL02}.
\textcolor{black}{The very existence of a flat one-magnon band is connected
with destructive quantum interference
which is related 
to the geometric structure of corner-sharing triangles. To get an imperession of the band-structure
of the one-magnon excitations above saturation we refer the reader to
Ref.~\cite{Mizoguchi2021},
where the band structure is  shown for a fermionic model
on the square-kagome lattice
which is closely related in terms of the one-particle spectrum. }

Similar to the  kagome HAF, see \cite{NSH:NC13,CDH:PRB13,Rausch2022},   
the plateau states of the spin-half SKHAF are
non-classical valence bond states. In the upper plateau it is the
exactly known magnon-crystal state built of one-magnon states on the squares
and $z$-aligned spins on intermediate A sites on the connecting triangles
\cite{SSR:EPJB01,richter2004-spin-peierls,richter2009squago}. The
lower $1/3$ plateau state 
is not exactly known. The inspection of the expectation values 
$\langle \op{s}^z_i \rangle$ and    
  $\langle  \op{\vec{s}}_i \cdot \op{\vec{s}}_j\rangle$
provides strong evidence  that the squares carry an almost perfect singlet
state and again the  spins on intermediate sites are $z$-aligned. We find that
the total spin on a square  $\langle\op{\vec{S}}^2_{\rm square}\rangle
=\sum_{i,j \in \rm square} \langle \op{\vec{s}}_i \cdot \op{\vec{s}}_j\rangle
= 0.06$ is close to zero and the local $z$-components of spins on 
A and B  sites  
$\langle \op{s}^z_{i \in \rm A} \rangle= 0.454$ and $\langle \op{s}^z_{i \in
\rm B} \rangle= 0.023$.

We present the influence  of the temperature on the magnetization curve in
Fig.~\ref{fig_M_H_T}. 
It is obvious that  for
elevated temperatures the experimental detection of plateaus becomes
difficult, particularly, if the plateau width is only of moderate size, see
the upper part of the $M(B,T)$ curve.
Moreover, we notice
that the jump of the magnetization to saturation at $T=0$ is washed out
already for small temperatures.

To detect plateaus and jumps in experiments   
the first
derivative  $dM/dB$ as a function of $T$ is more suitable, cf., e.g., Ref.~\cite{Tanaka_trian_2012}.
We show $dM/dB$ for $N=42$ in Fig.~\ref{fig_dM_dH_T}. 
We notice first that the oscillations present for the lowest temperature $T=0.03$ (red
curve)
are finite-size effects.
The plateaus at $1/3$ and $2/3$ show up as pronounced minima  in
$dM/dB$, however, the detection of such minima requires sufficiently low
temperatures.  
The jump of the magnetization to saturation at $T=0$ (washed out in the $M(B)$
curve at $T>0$)
leads to a high peak in $dM/dB$ at the saturation field.
The smaller peak at the upper end of the $1/3$ plateau is related to the
enlarged step (twice as high as the normal finite-size step). 
Contrary, to the jump to saturation we may expect that it will disappear as $N
\to \infty$. 
The melting of the pronounced plateaus at $1/3$ and $2/3$
with growing temperature happens more rapidly for the upper 2/3 plateau.
Furthermore, the 2/3 plateau melts asymmetricly, i.e. the
minimum in $dM/dB$ is below the midpoint of the plateau.

The influence of the magnetic field $B$ on the specific heat $C$ for $N=42$ is depicted in
Fig.~\ref{fig_C_var_H}(a) for selected $B$ values as shown in the inset of
Fig.~\ref{fig_C_var_H}(a).
Below the $1/3$ plateau and at very low temperatures 
the influence of
$B$ is determined by the shift of the low-lying magnetic excitations
with  $M=1,2$ and $3$ towards and even beyond the zero-field
singlet ground state. As a result, the low-$T$ shoulder [present for $B=0$, line 1
in Fig.~\ref{fig_C_var_H}(a)] may become
a low-temperature peak in
$C(T)$ (line 2), where its position  and height depends on $B$.
A similar low-$T$ scenario is observed  for $B$ between the  $1/3$ and the
$2/3$ plateaus (line 4) and for $B$ above the $2/3$ plateau (line 6).
A striking low-$T$ feature for $B$-values inside the plateaus (lines 3 and
5), where the ground state is a valence-bond state (see above),  is the well-pronounced
pretty high extra peak.
The extra peak for $B$ within the $2/3$ plateau is well-understood. It is a
flat-band effect and is related 
to the huge manifold of low-lying localized multi-magnon
states
\cite{zhitomirsky2004exact,Derzhko2006universal,DRH:LTP07,DRM:IJMP15,kago42,PRL_mag_cryst}.
Apparently, it is size-independent (see Fig.~\ref{fig_C_var_H}(b)), i.e., it 
persists for $N\to\infty$.
It is worth mentioning the contrast to the kagome HAF which is related to
the different structure of the respective  localized-magnon states. While these
states for the SKHAF are located on isolated trapping cells (squares), 
the trapping cells (hexagons) of the kagome HAF are connect leading to a
repulsive interaction of the traps.
As a result, this
extra maximum of the specific heat of the kagome HAF increases with $N$ and in the thermodynamic limit
it  becomes a true singularity indicating a
low-temperature order-disorder
transition into a magnon-crystal
phase \cite{zhitomirsky2004exact,PRL_mag_cryst}.
For the extra peak when $B$ is inside the $1/3$ plateau
there is no  straightforward
explanation. Our data for $N=30,36,42$ shown in Fig.~\ref{fig_C_var_H}(b)
indicate that there is only a small finite-size effect and the height of the
maximum is slightly growing with $N$. We may conclude that most likely it also
persists for
$N\to\infty$.

Let us turn to the main maximum around $T \sim 1$, see
Fig.~\ref{fig_C_var_H}(a).  At a first glance, the variation of
this maximum around $T \sim J$ seems
to be not very systematic.
However, as discussed for the kagome
HAF \cite{kago42} the effect of $B$ on the main maximum of $C$
is influenced by the huge manifold of localized-magnon (flat-band) states.
We compare  the height 
$C_{\rm max}$ and the position of the main maximum $T_{\rm max}$ for the
$N=42$ SKHAF, the $N=42$  kagome HAF and the  $N=40$ square-lattice HAF
in Fig.~\ref{fig_Cmax_Tmax}.
At low magnetic fields the value of $T_{\rm max}$
is determined by $J$ and the coordination number $z$ and therefore the
behavior of  $T_{\rm max}$
is very similar for all models.
While  $C_{\rm max}$
remains almost constant until  $B \sim 0.8
B_{\rm sat}$ and then it increases
smoothly for  $B >B_{\rm  sat}$ the variation of $T_{\rm max}$ as a function of
$B$ is more pronounced. It exhibits two maxima at $B=0$ and $B \approx 1.1 B_{\rm
  sat}$ and two minima at   $B \approx 0.5 B_{\rm  sat}$ and
$B\approx 1.4 B_{\rm  sat}$ as well as two regions 
with an approximately  linear increase of $T_{\rm max}$.
Again, the remarkable agreement of the SKHAF and the kagome HAF is obvious.
On the other hand, the difference to the square-lattice HAF in  $C_{\rm max}$ and more
pronounced in $T_{\rm max}$ is striking.
While the difference in $C_{\rm max}$ at low $T$ is 
related to the different low-energy physics which affects $C$ at higher
$T$ according to the sum rule $\int_0^\infty \frac{C(T)}{N T} dT =\int_{T=0}^{T=\infty}\frac{1}{N} dS =
k_B \ln(2)$, 
the different behavior of  $T_{\rm max}(B)$   beyond $B \sim 0.5 B_{\rm  sat}$
signals flat-band effects, i.e. the exponentially growing number of localized multi-magnon states
\cite{Derzhko2006universal} yields an increase of  $T_{\rm max}$
up to a  noticeable maximum  around  $B = B_{\rm  sat}$.
It follows a linear increase of $T_{\rm max}$ above $B_{\rm  sat}$
present in all models which is linked to the paramagnetic
phase.

%%%%%%%%%%%%%%%%%%%%%%%%%%%%%%%%%%%%%%%%%%%%%%%%%%%%%%%%%%%%%%%%%%%%%%%%
\section{Conclusions}
\label{sec-4}

In our study
 we performed  large-scale calculations of
thermodynamic quantities such as the magnetization $M(T)$, the specific
heat $C(T)$, the entropy $S(T)$ and the susceptibiliy $\chi(T)$ for the highly frustrated spin-half square-kagome Heisenberg antiferromagnet
(SKHAF).
For that we used the finite-temperature Lanczos method (FTLM) applied to
seven different finite lattices including the 'large' lattices  of $N=42, 48$
and $54$ sites, where for $N=48$ and $54$ only the thermodynamics  near
the saturation field was considered.
At zero magnetic field,  we find a remarkable accordance of the thermodynamic
properties
 of the SKHAF
with those of the paradigmatic kagome HAF. 
Our 
results for $N=42$
and smaller sizes indicate that the specific heat very likely has
got a low-temperature shoulder instead of an additional
low-temperature maximum. 
There is also a considerable influence of frustration on the susceptibility
and
on the entropy. Thus we find 
a noticeable shift of the maximum of $\chi(T)$ to low $T$ compared to
two-dimensional unfrustrated HAFs, and, the entropy per site $S(T)/N$ acquires
about
$40$\% of its  maximum value $\ln2$ already  at $T/J \sim 0.1$.

Other interesting properties  of the SKHAF are related to the magnetization
process in an applied magnetic field $B$.
There are two well pronounced plateaus at $1/3$ and $2/3$ of the saturation
magnetization and a jump from the $2/3$ plateau directly to saturation caused
by the flat one-magnon band.
The 
melting of the  plateaus with growing temperature is
faster for the upper  $2/3$  plateau and it 
melts asymmetricly.
For magnetic fields inside the plateaus the specific heat 
shows a well-pronounced
extra peak at low temperatures, which exhibits only small finite-size
effects.
Furthermore, we find that the influence  of strong frustration is not
only visible at low 
temperatures $T \ll J$, it is also noticeable at moderate (and high)
temperatures $ T \sim J$.
Especially, the presence of a flat one-magnon band leading to
a huge manifold of low-lying flat-band states
yields  pronounced effects in the magnetization process and
the temperature dependence of the specific heat  at magnetic fields
above $B \sim 0.5B_{\rm sat}$. 

Though our investigation  of the spin-$1/2$ SKHAF as a highly frustrated quantum spin system is of interest 
in its own right, it is also motivated 
by the recent discovery of
a spin liquid in the square-kagome magnet
KCu$_6$AlBiO$_4$(SO$_4$)$_5$Cl \cite{FMM:NC20}, which exhibits, however,
three different exchange couplings.
Moreover,
the large variety of magnetic insulators \cite{Ino:AP18,QSL_Materials_2021}
as well as the progress in synthesizing new magnetic molecules and compounds with predefined   
spin lattices may open the window to get access to the observation of the
discussed phenomena.

Bearing in mind the numerous studies of the low-energy physics of the
related kagome HAF 
we argue that our work may also stimulate other studies using
alternative techniques, such as tensor network
methods, DMRG, numerical linked cluster expansion or Green's function techniques
\cite{Pollmann2017,Xi-Chen2018,SHM:PRB20,kagome-RGM2018,Rausch2021}.

%%%%%%%%%%%%%%%%%%%%%%%%%%%%%%%%%%%%%%%%%%%%%%%%%%%%%%%%%%%%%%%%%%%%%%%%
\section*{Acknowledgment}

This work was supported by the Deutsche
Forschungsgemeinschaft (DFG  RI 615/25-1 and SCHN 615/28-1). Computing time at
the Leibniz Center in Garching is gratefully acknowledged.

%%%%%%%%%%%%%%%%%%%%%%%%%%%%%%%%%%%%%%%%%%%%%%%%%%%%%%%%%%%%%%%%%%%%%%%%
\appendix
\section{Finite square-kagome lattices used for 
the exact diagonalization 
and the
finite-temperature Lanczos method}
\label{sec-a}

\begin{figure}[ht!]
\centering
\includegraphics*[clip,width=0.95\columnwidth]{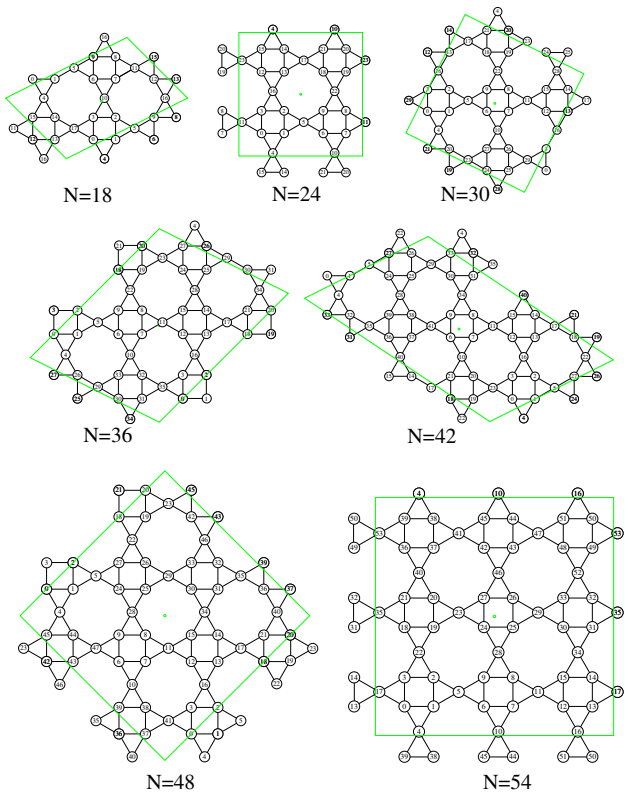}
\caption{Finite lattices used for FTLM.
The structures 24, 30, 36, 42 are the same as used in
\cite{Sakai2015}.}
\label{fig_lat}
\end{figure}

Here, we provide the employed lattice structures in Fig.~\ref{fig_lat}.

%===================    figure   =================================
%===================    figure  =================================

%%%%%%%%%%%%%%%%%%%%%%%%%%%%%%%%%%%%%%%%%%%%%%%%%%%%%%%%%%%%%%%%%%%%%%%%
\bibliography{JR_RGM,mag_plateaux,JR_own,js-own,js-other,allerlei,sawtooth}

\begin{thebibliography}{73}
\bibitem [{def()}]{def_Archi}%
  \BibitemOpen
  \href@noop {} {}\bibinfo {note} {An {Archimedean} lattice is built by uniform
  polygons. Importantly, it has only one set of topologically equivalent
  sites.}\BibitemShut {Stop}%
\end{thebibliography}

\end{document}